\documentstyle[aps,preprint]{revtex}
\begin{document}
\draft

\title{Correlated few-electron states in vertical double
quantum dot systems}
\author{Juan Jos\'e Palacios$^*$}
\address{Departamento de F\'{\i}sica de la Materia Condensada, Universidad
Aut\'onoma de Madrid, Cantoblanco 28049, Madrid, Spain.}
\author{Pawel Hawrylak}
\address{Institute for Microstructural Sciences, National Research Council of
Canada, Ottawa, Canada K1A 0R6.}

\maketitle

\begin{abstract}
The electronic properties of semiconductor, vertical,
double quantum dot systems with few electrons are
investigated  by means of analytic,
configuration interaction, and mean field
methods. The combined effect of a high magnetic field,
electrostatic confinement, and inter-dot coupling,
induces a new class of few-electron ground states absent in
single quantum dots. In particular, the role played by
the isospin (or quantum dot index) in determining the appearance of
new ground states is analyzed and
compared with the role played by the standard spin.
\end{abstract}

\date{}

\section{Introduction}

The  behavior of a small number of electrons confined into a single
quasi-two-dimensional
quantum dot (QD) in the presence of a
magnetic field has been studied over the past few
years.$^{\cite{bryant,cha1,merkt,john,pawel,eri,pal,palacios}}$
Most of the work has focused on
the high magnetic field regime of fully spin-polarized electrons  where
incompressible electron states analogous to the Laughlin
states of the fractional quantum Hall effect exist.
Only very recently
the problem of spin and its implications  has been
addressed.$^{\cite{eri,pal,palacios}}$ The ground and excited
states of a spin-unpolarized QD turn out to be  much more
complex than those of the polarized QD.
The role of spin and spin induced interactions can be effectively
simulated by the {\em isospin} (layer index) in double layer systems where the
isospin-up isospin-down states correspond to electrons
on layer 1 or layer 2. Recent
experimental$^{\cite{suen1,eisen,mur,suen2}}$ and
theoretical$^{\cite{halp,rez,cha2,yos,fer,mac1,mac2,brey,he,cote,mac3,varma}}$
studies of a double layer two-dimensional electron
system (DL2DES) in a high magnetic field
have shown a rich variety of new incompressible states related
to the quantum Hall effect$^{\cite{pra}}$ (QHE). For instance,
in recent experiment
of Ref.\ 23 S. Q. Murphy {\em et al.} have reported the existence
of a new incompressible QHE state at a filling factor of individual layers
$\nu=1/2$$^{\cite{pra1}}$ in the absence of inter-layer tunneling.
This coupling-induced QHE
state has no analogue in single layer systems. The distance
between layers, i.e., the strength of the Coulomb coupling, determines whether
the incompressible state appears or not. This experiment seems to
confirm the predictions made in Refs.\ 15-17, and 21 that the new QHE states
are supported by the inter-layer Coulomb interactions and by
the single-particle symmetric-antisymmetric gap in the presence of arbitrary
tunneling. The breakdown of these states occurs
when the Coulomb coupling is switched off and the intra-layer correlations
become dominant.$^{\cite{mac1,mac2,brey,he,cote}}$

In this work we examine the role of confinement, magnetic field, and inter-dot
coupling
in the generation of a new type of few-electron states
in a vertical double quantum dot system (DQDS). We concentrate on
the high magnetic field limit where electrons in the ground and low-lying
states are spin polarized due to Zeeman
energy, and the effect of the isospin degree of freedom can be isolated.
the appearance of several ground states (GS) with an unexpected isospin
is the most striking feature
of such systems in comparison with isolated polarized QD's .

The paper is organized as follows: Section I is devoted to the description
of the model; in Sec. II we present the analytical results for the simplest
possible
case of two electrons in a DQDS; in Sec. III we present numerical results for
up to six electrons, and Section IV contains conclusions of the work.

\section{The model}

We consider a pair of identical vertical
coupled QD's with
 parabolic in-plane confinement, containing $N$ electrons, and
separated by a distance $D$ as shown in Fig.\ \ref{fig1}. The normalized
single-particle states $|m,n;\sigma>$ of each individual QD
(in the presence of
a magnetic field $B$ oriented normal to the plane of the QD) are simply
harmonic oscillator states

\begin{equation}
|m,n;\sigma>=\sqrt{\frac{1}{m!n!}} (a^\dagger)^m (b^\dagger)^n|0,0;\sigma>
\end{equation}

with $\sigma$ labeling the QD index or $z$-component of the
isospin quantum number.
 In analogy with the standard spin,  $\sigma$ takes on the values
$+1/2,-1/2$  (for brevity, $+,-$, from now on)
for electrons in the "upper" and "lower" QD, respectively.
The single-particle energies are those of a pair of harmonic oscillators
(the Zeeman energy is omitted
and  $\hbar=1$ for the rest of this work)

\begin{equation}
\epsilon_{mn\sigma}=\Omega_+(n+\frac{1}{2})+ \Omega_-(m+\frac{1}{2})
\end{equation}

with $\Omega_{\pm}= [\sqrt{\omega_c^2+4\omega_0^2} \pm\omega_c]/2$.
$\omega_c$ is the cyclotron frequency and $\omega_0$ is the frequency
characterizing the parabolic confinement of both QD's.
We shall restrict to the case
of high magnetic fields ($B>2-3$ Tesla) where electrons are spin polarized,
and the condition $\omega_c > \omega_0$ is satisfied. In this regime
one can restrict the basis set to $n=0$ and the problem
becomes one-dimensional. We shall omit index $n$ in what follows.

Allowing for inter-dot tunneling, the full-interacting
Hamiltonian of the DQDS can be expressed in second quantization as

\begin{eqnarray}
H&=&\sum_{m\sigma\sigma'} [\delta_{\sigma,\sigma'}
\epsilon_{m\sigma} c^\dagger_{m\sigma} c_{m\sigma'}
+ (1-\delta_{\sigma,\sigma'}) t
 c^\dagger_{m\sigma} c_{m\sigma'}] + \nonumber \\
& & \frac{1}{2}
\sum_{m_1m_2m_3m_4\sigma\sigma'} V^{m_1m_2m_3m_4}_{\sigma\sigma'}
c^\dagger_{m_1\sigma}c^\dagger_{m_2\sigma'} c_{m_3\sigma'} c_{m_4\sigma},
\label{ham}
\end{eqnarray}
where $c^+$, $c$ are the single-particle creation and
annihilation operators,  $ V^{m_1m_2m_3m_4}_{\sigma\sigma'}$ the Coulomb
interaction terms, and $t$ is the inter-dot hopping matrix element.
We have included here inter-dot direct and exchange terms but neglected some
off-diagonal scattering elements, negligible  in the
weak hopping ($t \rightarrow 0$) limit.

Alternatively, we can use an isospin 1/2 operator representation to
define isospin operators in terms of previous
creation and annihilation single-particle operators as
\begin{eqnarray}
\rho_{m_1m_2} & = & c^\dagger_{m_1+} c_{m_2+} +
 c^\dagger_{m_1-} c_{m_2-} \nonumber \\
\zeta_{m_1m_2}^z & = & c^\dagger_{m_1+} c_{m_2+} -
 c^\dagger_{m_1-} c_{m_2-}\nonumber  \\
\zeta_{m_1m_2}^x & = & c^\dagger_{m_1+} c_{m_2-} +
  c^\dagger_{m_1-} c_{m_2+}.\nonumber
\end{eqnarray}

Omitting linear terms in $\rho_{m_1m_2}$ and $ \zeta_{m_2m_3}^z$
allows us to emphasize the appearance of new
 isospin-isospin interactions in the DLQD Hamiltonian:
\begin{eqnarray}
H& =&\sum_m (\epsilon_m \rho_{mm} + t \zeta^x_{mm}) +
\frac{1}{2}
\sum_{m_1m_2m_3m_4} V^{m_1m_2m_3m_4}_D \rho_{m_1m_4}
 \rho_{m_2m_3} \nonumber \\
& &+\frac{1}{2} \sum_{m_1m_2m_3m_4} V^{m_1m_2m_3m_4}_E
\zeta_{m_1m_4}^z  \zeta_{m_2m_3}^z.
\label{hamisos}
\end{eqnarray}
The Hamiltonian (\ref{hamisos})
contains two $SU(2)$ symmetry-breaking fields: The first one along the $x$
direction and proportional to the hopping $t$,
and the second one along the $z$ direction,
the isospin-isospin interaction, proportional to $V_E$
(in addition to the usual charge-charge
interaction). The Coulomb matrix elements are given by
\begin{eqnarray}
V^{m_1m_2m_3m_4}_D & =& \frac{1}{2}(V_{++}^{m_1m_2m_3m_4}+
V_{+-}^{m_1m_2m_3m_4}) \nonumber \\
V^{m_1m_2m_3m_4}_E & = & \frac{1}{2}(V_{++}^{m_1m_2m_3m_4}-
V_{+-}^{m_1m_2m_3m_4}),  \nonumber
\end{eqnarray}
where $V_{++}$ (identical to $V_{--}$)
are intra-dot Coulomb matrix elements, and $V_{+-}$
are inter-dot Coulomb matrix elements.

As far as the symmetry-breaking term $t$ is concerned, two physical
situations can be distinguished: the "incoherent" and the "coherent" one.
 The incoherent case ($t=0$) describes isolated QD's coupled only
by Coulomb interactions. Electrons cannot transfer (exchange) between QD's,
and are localized on individual QD's. This distinguishability manifests itself
in the anticommutation relation $\{c^+_{m_1\sigma},c_{m_2\sigma'}\}= \delta_{
\sigma,\sigma'}\delta_{m_1,m_2}$. This is the well known "layered electron
gas" model.$^{\cite{pawel1}}$

The second case is the coherent behavior of QD's where electrons cannot be
in a specific QD but occupy the symmetric ($s$)
or antisymmetric ($as$) orbitals of a pair
of QD's. The transformation from orbitals localized on individual dots
to the symmetric/antisymmetric orbitals is equivalent to the
rotation of the isospin. Let us define a rotated isospin representation,
 $\{\alpha\}$, which
diagonalizes the hopping part of the Hamiltonian (\ref{hamisos}) for arbitrary
hopping matrix element $t$:
\begin{eqnarray}
c^\dagger_{m+}&=&\frac{1}{\sqrt{2}}
(\alpha^\dagger_{m,s}+\alpha^\dagger_{m,as})\nonumber \\
c^\dagger_{m-}&=&\frac{1}{\sqrt{2}}
(\alpha^\dagger_{m,s}-\alpha^\dagger_{m,as})\nonumber .\\
\end{eqnarray}
We wish to emphasize that this transformation does not depend on the
strength of the hopping matrix element.
We can define isospin operators $\rho,\zeta^z,\zeta^x$ in the space of
coherent operators $\{\alpha\}$ to write the coherent
Hamiltonian as
\begin{eqnarray}
H&=&\sum_m (\epsilon_m \rho_{mm} + t \zeta^z_{mm}) +
\frac{1}{2}
\sum_{m_1m_2m_3m_4} V^{m_1m_2m_3m_4}_D \rho_{m_1m_4}  \rho_{m_2m_3}\nonumber \\
&&+\frac{1}{2} \sum_{m_1m_2m_3m_4} V^{m_1m_2m_3m_4}_E
\zeta_{m_1m_4}^x  \zeta_{m_2m_3}^x.
\label{hamcoh}
\end{eqnarray}
In this rotated isospin space the hopping matrix element $t$
is simply equivalent to an
external field. The coherent Hamiltonian (\ref{hamcoh})
is similar to the previous one  (\ref{hamisos}) in the sense that it also
presents two $SU(2)$ symmetry-breaking terms: that proportional
to the hopping $t$ and the isospin-isospin
interactions proportional to $V_E$
(again, in additon to the normal $SU(2)$ invariant charge-charge interactions).
This isospin-isospin interaction $V_E$,
present in both Hamiltonians, is, ultimately,
responsible for the novel physics in
a DQDS. Since this physics is dominated by Coulomb interaction
we shall concentrate on the $t\rightarrow 0$ limit.

\section{Analytical results for the two-electron DQDS}

The case of $N=2$ is the simplest case that
deserves to be studied in detail and can be solved
analytically.$^{\cite{neil}}$
We can write the total wave function $\Psi(\stackrel{\rightarrow}{r_1},{z_1}
\stackrel{\rightarrow}{r_2},{z_2})$ as a product
of the center of mass ($\stackrel{\rightarrow}{R}$)
wave function, relative motion ($\stackrel{\rightarrow}{r}$) wave function,
and the rotated isospin wave functions (symmetric or antisymmetric)
\begin{equation}
\Psi(r_1,r_2)=\psi^{M^{cm}}(\stackrel{\rightarrow}{R})
\phi^{M^r}( \stackrel{\rightarrow}{r})
|i,j> \equiv |M^{cm},M^r;i,j> \;\;\;\;\;\;i,j=s,as.
\end{equation}
The center of mass motion (characterized by an angular momentum $M^{cm}$)
separates from the Hamiltonian, and isospin states can be characterized by
 4 orthogonal rotated isospin states \{$|s,s>,|as,as>,
|s,as>,|as,s>$\}. These states can be written in a more familiar language
of the usual isospin states, e.g., \{$|+,+>=|I=1,I_z=+1>$\},
based on orbitals localized on individual QD's:
\begin{equation}
\begin{array}{lll}
|s,s>&=&\frac{1}{\sqrt{2}}\left\{\frac{|I=1,I_z=+1>+|I=1,I_z=-1>}{\sqrt{2}}
+|I=1,I_z=0>\right\}  \\
|as,as>&=&\frac{1}{\sqrt{2}}\left\{\frac{|I=1,I_z=+1>+|I=1,I_z=-1>}{\sqrt{2}}
-|I=1,I_z=0>\right\}  \\
|s,as>&=&\frac{1}{\sqrt{2}}\left\{|I=1,I_z=+1>-|I=1,I_z=-1>\right\} \\
|as,s>&=&|I=0,I_z=0>.
\end{array}
\end{equation}
The first three states correspond to a well-defined isospin $I=1$
but an undefined $z$ component of $I$. The expectation value of $I_z$
for the three coherent states $I=1$ is  zero, but quantum fluctuations
are  present in the $I=1$ states, in contrast with the incoherent $I=0$ state.
This means that the two $I=1$ states $\frac{1}{\sqrt{2}}\{|s,s>+|as,as>\}$
and $|s,as>$
are not eigenstates of $I_z$, and one cannot determine
on which QD the electrons are localized.
They correspond to having both electrons in one or the other QD of the
DQDS.  On the other hand, the $|I=1,I_z=0>$ and $|I=0,I_z=0>$ states
($\frac{1}{\sqrt{2}}\{|s,s>-|as,as>\}$ and $|as,s>$, respectively)
correspond to electrons on opposite QD's and their energy is determined
only by inter-dot interaction $V_{+-}=V_D-V_E$.

We can now write the relative particle Hamiltonian  in the rotated
isospin space as a $4\times 4$ matrix. It is easy to see that the first three
states correspond
to total isospin $I=1$ and hence the in-plane relative particle wave function
must be
antisymmetric. This corresponds to {\em odd} relative angular momentum,
$M^r$, of the relative particle. The Hamiltonian $H_1$
for the $I=1$ states can be written as
\newpage
\begin{eqnarray}
&&\left( \begin{array}{ccc}
t+\Omega_-M^r+<M^r|V_D|M^r>& <M^r|V_E|M^r>& 0 \\
 <M^r|V_E|M^r>& -t+\Omega_-M^r+<M^r|V_D|M^r> & 0  \\
0 & 0& \Omega_-M^r+<M^r|V_D+V_E|M^r> \end{array} \right), \nonumber \\
&& \nonumber \\
\label{h1}
\end{eqnarray}
and for $I=0$ we simply have
\begin{eqnarray}
\Omega_-M^r+<M^r|V_D-V_E|M^r> = H_0,
\end{eqnarray}
with relative angular momentum $M^r$ being {\em even}.

{}From the rotated isospin Hamiltonian (\ref{h1}) it is clear that
(a) only the symmetric-antisymmetric states
are coupled,
(b) coupling is due to the symmetry-breaking exchange interaction $V_E$,
(c) the coupling between symmetric-antisymmetric states
is present even in the absence of tunneling ($t=0$), and
(d) only these coupled symmetric-antisymmetric states are affected by
inter-dot hopping ($t$).

The two-electron Hamiltonian can be
easily diagonalized and simple analytical expressions for energies and
wave functions obtained (not shown here for brevity).
Figure \ref{fig2} shows the evolution of the total energy
spectrum as a function of the distance between QD's
for a given value of the total angular momentum $M=M^r+M^{cm}=6$ and
zero hopping matrix element $t$.
Standard values of dielectric constant and effective mass
for GaAs have been taken throughout the calculations and the
confining energy of the QD's in the DQDS is taken to be 5 meV. From now on
it is also convinient to define an inter-dot coupling constant:
$\alpha=V^{0000}_{+,-}/V^{0000}_{+,+}$.
One can see in Fig.\ \ref{fig2} the energy splitting of
the $I=1$ states (when $M^r$ odd), which
are degenerate when $\alpha=1$, as the distance increases
(i.e., as the coupling constant $\alpha$ lowers its value).
Those with $I_z=0$ go down in energy while those
with $I_z=\pm 1$ remain degenerate and constant (notice that this would not
be the case any more if $t \neq 0$). It must be pointed out that the total
isospin $I$ is a good quantum number for any value of the coupling
constant in the two-electron DQDS.
This is no longer the case for a higher number of electrons as
will be shown below.
In Fig.\ \ref{fig3} we show the phase diagram  (isospin $I$ and angular
momentum $M$) of the two-electron DQDS
GS as a function of the magnetic field $B$ and coupling constant $\alpha$.
The magnetic field changes the ratio of kinetic ($\Omega_-$) to Coulomb energy
and induces changes in the isospin and angular momentum
of the DQDS GS. Similar transitions of the
isospin and angular momentum of the GS can be induced by changing the
coupling constant as shown in Fig.\ \ref{fig3}.

\section{More than 2 electrons in the DQDS}

\subsection{Zero distance limit: No SU(2) symmetry-breaking interactions}

The zero distance limit (ZDL) ($D=0$, $\alpha=1$), i.e., the
limit of having the QD's forming the DQDS superimposed in real space,
although urealistic, is interesting and deserves to be studied in detail.
In this particular case (always with $t\rightarrow 0$)
the $SU(2)$ isospin symmetry is not broken by the
isospin-isospin interactions since $V_E=0$. In the ZDL the DQDS of
identical QD's under high magnetic fields
is completely equivalent to a single QD with  Land\'e factor $g$
$\rightarrow 0$, i.e., in the zero  Zeeman limit.
The role of spin in the DQDS
(frozen out by the magnetic field) is now  played by the isospin.
In the case of non-identical QD's, the presence of an "isospin Zeeman energy"
(for instance, a difference between the confinement energies of each QD)
would make the DQDS equivalent to the spin-polarized single QD .
The feasibility of fabricating identical QD's with zero
"isospin Zeeman energy"
is one of the most appealing possibilities
presented by such systems. As will be shown below, it is
the fundamental origin of the new electronic
properties that appear in a DQDS
at high magnetic fields, compared to those appearing in the same regime
of fields in a single QD (i.e., in a spin-polarized QD).

We now extend our study to a larger number of electrons by employing exact
diagonalization techniques.
The intra- and inter-dot electron-electron correlations have been
taken fully into account by expanding the many-body wave function
in terms of
Slater determinants (or configurations) and diagonalizing
the Hamiltonian (\ref{ham}) for 4 and 6 electrons.
The Hilbert space  has been restricted to the states of
the lowest Landau level  of each
QD ($n=0$). Due to the circular symmetry, the diagonalization can be
done in separate subspaces of configurations with the same
$z$-component of the total angular momentum, $M$.
In the absence of hopping any subspace of given $M$  can be split,
in turn, into orthogonal subspaces of given $I_z$. In this way the size of the
matrices to diagonalize becomes smaller and computationally more accesible.
The GS's and lowest-lying eigenvalues and eigenfunctions of all the
subspaces of different $M$
were obtained using standard diagonalization routines.

Figures \ref{fig4}(a) and (b)
show the evolution of $M$ for the DQDS absolute GS
as a function of $\Omega_-$ (i.e., $B$) for 4 and 6 electrons .
The absolute GS angular momentum $M$ goes through a series of
increasing values as $\Omega_-$ lowers its value (i.e,
the magnetic field rises). The competition between kinetic energy and Coulomb
repulsive energy determines the value of $M$:  the kinetic energy
(due to the confinement and magnetic field)
favors electrons in the center of
the QD, the Coulomb repulsion tends to spread the charge.

The results in Figs.\ \ref{fig4}(a) and (b) show a remarkable
stability
of the GS's with $M=6, I=2$ for 4 electrons, and $M=15, I=3$
for 6 electrons, against changes in $\Omega_-$.
These stable GS's are the only ones appearing with maximum
total isospin $I$, and
are five-fold degenerate ($I_z=2,1,0,-1,-2$) and seven-fold degenerate
($I_z=3,2,1,0,-1,-2,-3$), for 4 and 6 electrons, respectively.
In particular, the $I_z=0$ states are
described {\em exactly} in this ZDL by the Jastrow-type
correlated wave function [111]$^{\cite{halp}}$
\begin{eqnarray}
\Phi(z_1,...,z_{N/2},w_1,...,w_{N/2}) & = & A [
\prod_{1\leq i < j \leq N/2} (z_i-z_j)
\prod_{1\leq i < j \leq N/2} (w_i-w_j) \prod_{1 \leq i,j \leq N/2 }
(z_i-w_j) \times  \nonumber \\
& & \exp (-\sum_{i=1}^{N/2} |z_i|^2/4l^2 -
\sum_{i=1}^{N/2} |w_i|^2/4l^2) \times \nonumber \\
& &|+>_1...|+>_{N/2}|->_1...|->_{N/2}],
\end{eqnarray}
proposed by Halperin in the context of fractional QHE
wave functions with spin degrees of freedom, and used later in
the context of a DL2DES.$^{\cite{yos}}$
The electron coordinates of the upper dot
are given by $z_i=x_i-iy_i$ and those of the lower dot by $w_i=x_i'-iy_i'$,
and $+$ and $-$ denote the values of the QD indices (see previous section).
The symbol $A$ represents the antisymmetrization operator. Alternatively,
these states can be also expressed in terms of the
single-particle occupation numbers of each QD, $\nu^+$ and $\nu^-$.
For instance, for 6 electrons
$\nu^{+,-}_m=0.5$ for $m=0,1,2,3,4$, and 5, and 0 for the following $m$'s.
If we define a filling factor for the DQDS as
$\nu^{DQDS}=\langle \nu^+_m + \nu^-_m \rangle _m$ where the brackets
denote average over the lowest occupied $m$'s, the GS's $M=6$ ($N=4$)
and $M=15$ ($N=6$,) clearly correspond to $\nu^{DQDS}=1$.
These $\nu^{DQDS}=1$ states can
be considered, in turn, as precursors of the incompressible state at total
$\nu=1$ observed in a DL2DES.$^{ \cite{mur}}$

In Fig.\ \ref{fig5} we show the six-electron excitation spectrum of a
DQDS with a $\nu^{DQDS}=1$ degenerate GS of angular momentum $M=15$
(marked with a star in the plot). In particular,
the fully isospin-polarized state within the degenerate subspace
has been considered as the absolute GS in what follows.
The excitation spectrum consists of two branches: the branch with
$M<15$ and that with  $M>15$. Let us first
concentrate on the branch of the spectrum with angular momentum $M<15$.
By examining the wave function we have been
able to identify those excitations corresponding to
{\em isospin-flip} excitations (solid dots in the plot).
Those with $M$ close to $M=15$ can be associated to {\em
isospin-flip spin-wave-like edge} excitations.$^{ \cite{kallin}}$
Those with $M$ farther below
$M=15$ correspond to {\em isospin-flip quasiparticle} excitations
(magnetoexcitons$^{ \cite{kallin}}$) and
consist in flipping the electron's isospin and moving it
from the edge of the $\nu^{DQDS}=1$
droplet to a reversed isospin single-particle
state $m$ closer to the center of the
DQDS. The value of the total isospin of such excitations is
the maximum possible value according to the spin flip and to the
subspace in which it is found, i.e, the
value of $M$ of the excitation (all those shown with solid dots
in Fig.\ \ref{fig5}, on the left branch,
correspond to $I=(N-1)/2=5/2$). We can see that isospin-flip excitations are
{\em not}, in general, the lowest energy states in subspaces of given $M$,
but there appear a few other states of lower energy and minimum value
of $I$ within each subspace.$^{\cite{Rezayi}}$

We have also identified isospin-conserving
{\em quasihole-quasielectron pair}
excitations (open dots in the plot)
consisting of a quasihole in a single-particle state of $m<5$ and an electron
added to the edge of the droplet. Similarly to the spin-flip excitations,
those with $M$ far from $M=15$ present the better-defined character of a
single, localized hole in the droplet.
These excitations present the maximum
possible value of $I$ ($N/2=3$). In contrast with the
isospin-flip excitations, there appear {\em many}
excitations of lower energy and minimum isospin within each subspace of
given $M$.

{}From the excitation spectrum we can understand the
evolution of the absolute GS angular momentum  with $\Omega_-$.
The stability of the $\nu^{DQDS}=1$
GS's is nothing but the direct consequence
of the cusp-like structure or gap exhibited by the
excitation spectrum (see Fig.\ \ref{fig5}),
and such state can be referred to
as an {\em incompressible} GS. Up to the values of $\Omega_-$
studied, no other stable GS's seem to appear in our calculations.
In the case of identical QD's almost all the
lowest states of different subspaces  of $M$
will become the absolute GS of the DQDS at a certain value of $B$,
which is in clear contrast with the case of a single, isolated polarized QD.
In this ZDL the GS takes on almost all possible values of $M$ for the range
of variation of $\Omega_-$, but such values change very
quickly with $\Omega_-$
(except the $\nu^{DQDS}=1$), presenting no stability.
The relevance of these kind of
new absolute GS's in a single QD at low magnetic fields (when
the Zeeman energy cannot be considered infinite)
has been stressed in Ref.\ 8. Their importance lies
in suppressing, through the spectral function of the system,
the single-electron tunneling rates,
and therewith, in strongly modifying
the transport properties of the system. One
would expect that the presence of such states in the DQDS would give
rise to similar effects on transport properties of these systems
as long as they do not
disappear for a realistic situation, i.e, for a certain distance
between QD's. Taking into account realistic values of the distance
is the topic of the following sections.

As mentioned previously, a significant
difference between the confining energies of the
QD's forming the DQDS (or the presence of the $g$ factor for a single QD
at high $B$) tends to favor fully isospin-polarized (spin-polarized for
a single QD) absolute GS's.
QD's with fully polarized electrons have been extensively studied over the
past few years. It is known that, for instance, for
four and six electrons, when only the spin-polarized states
are relevant to the GS properties, the value of $M$ for the absolute GS
is restricted to a series of specific numbers: 6, 10, 14, ..., for 4
electrons$^{ \cite{trug,cha1}}$
and 15, 21, 25, ..., for 6 electrons$^{\cite{john}}$. These
numbers are known in the literature as "magic" numbers.$^{ \cite{cha1,pawel}}$
This can be seen easily from Fig.\ \ref{fig5}:
The lowest states in each subspace of
$M$ will correspond now (in the presence of isospin Zeeman energy)
to isospin-conserving quasihole
($M>15$) or isospin-flip quasiparticle ($M<15$) excitations. Only
those with quasiholes (quasiparticle)
close to the center of the droplet will become the
GS of the DQDS as $B$ changes [those with downward-pointing arrows in
Fig.\ \ref{fig5}].
A simple description of the magic absolute GS's (for $M>15$) in terms
of "bosonic" operators acting upon the electronic $\nu=1$ droplet
has been presented elsewhere.$^{ \cite{jacobo}}$

\subsection{Short distances: Weak SU(2) symmetry-breaking interactions}

We now discussed how the new minimum-isospin GS's of a ZDL DQDS
evolve for realistic distances between QD's.
Figures \ref{fig6}(a) and (b) show, for $D=100 \AA$ and $D=50 \AA$, the
evolution  with $\Omega_-$ of the $M$ value of the absolute
GS (from now on we will restrict
to the 6 electrons case). As was clearly shown in
Fig.\ \ref{fig2}, as the distance between QD's
increases, or the symmetry-breaking interaction term $V_E$ is stronger,
the isospin multiplet
degeneracy is removed and the GS will always have $I_z=0$,
i.e., equal number of electrons in each dot.
The total isospin, $I$, is no longer a good quantum number for $D \neq 0$,
but one can still trace it back to its original value at $D=0$, and
use it to label the states as long as the distance is not too large.
As can be seen in Figs.\ \ref{fig6}(a) and (b), for short distances
($D=50 \AA, \alpha \approx 0.8$) many of the absolute minimum-isospin
GS's in the ZDL survive.
The $M=15$ GS also remains stable. On increasing the distance ($D=100 \AA,
\alpha \approx 0.6$) the situation changes dramatically:
The symmetry-breaking interactions have made all the
minimum-isospin GS's of the ZDL disappear.
Instead, many new stable states appear ($M=6,9,12,15,18,...$),
and the only stable GS in the
ZDL, $M=15$, has become less relevant. At the same time, the overlap
of such GS  (or the Jastrow-type wave function shown above)
with those of the same $M$ at distances different from zero decreases with $D$.
Instead of showing this overlap we have chosen to show in
Fig.\ \ref{fig7} the single-particle
occupation numbers $\nu^+$ ($\nu^-$) in the ZDL, for $D=50 \AA$, and
for $D=100 \AA$. One can see
that the regular occupations forming the $\nu^{DQDS}=1$ state in the ZDL
melt as the distance between the QD's increases, and
the value of $M=15$  becomes less relevant for the GS.
These new stable states have their origin in the
"superposition" of the stable ones for each
QD with 3 electrons ($M=3,6,9,...$).$^{\cite{cha1,pawel}}$
The next section will clarify what we mean by such superposition.

\subsection{Large distance limit: Strong SU(2) symmetry-breaking interactions}

We have seen in the previous section that,
as the distance $D$ between dots increases the situation seems to change
noticeably. In order for this large-$D$ limit to be understood we have
carried out a a self-consistent Hartree-Fock (HF)
treatment of the inter-dot coupling, but conserving the intra-dot
correlation.
The procedure is the following. A  GS solution of total angular momentum,
$M^+$,
by means of an exact diagonalization of, for instance, the full interacting
Hamiltonian of the upper QD, $H^+$, is found as described in the previous
section.
Then, the single-particle energies of the lower dot are modified by

\begin{equation}
\epsilon'^-_m = \epsilon^-_m + \Sigma^+_m
\end{equation}

where, in the calculation of the self-energy $ \Sigma^+_m $,
Hartree-Fock-like diagrams have been used. If the inter-dot hopping
is forbidden, then only a Hartree-type diagram is allowed and corresponds to
taking into account the single-particle occupation numbers of the upper GS,
$\nu^+_m$,
together with the coupling term $V^{m_1m_2m_2m_1}_{+,-}$.
Now, an exact diagonalization
of the lower full interacting Hamiltonian, $H^-$,
with the corrected single-particle energies is done, a lower GS
of $M^-$ is found,
and the occupations of this GS are used to modify, in turn, the
single-particle energies of the upper dot. The process continues until
convergence is achieved.

Figure \ref{fig8} shows the evolution of the renormalized
GS energy of our DQDS as a
function of the magnetic field (or $\Omega_-$)
for the case of 3 electrons in each dot
($I_z=0$) and for four different distances. The GS's for
$D = 200 \AA$ ($\alpha \approx 0.3$) correspond to a superposition of
$M^+=3$, $M^-=3$ (which gives us
$M=M^++M^-=6$), $M^+=6$, $M^-=6$ ($M=12$), and $M^+=9$, $M^-=9$ ($M=18$).
The possible values for $M^+$ and $M^-$ (3,6,9,...)
are the corresponding magic angular momenta $M$
mentioned above for the case of 3 electrons in a single QD.$^{
 \cite{cha1,pawel}}$ As we bring the dots
together, these states shift to lower fields,
so one can think of the
inter-dot coupling in terms of an "additional" magnetic field
which adds to the bare one to give a stronger effective value of $B$.

Exact energies are also shown in Fig.\ \ref{fig8}.
At large distances ($D=200 \AA$) both procedures give the same energy for the
GS.  As the dots are brought together,
the exact GS energy becomes smaller than the HF energy.
This fact is more noticeably for the shortest distances where this deviation
takes place at  $B \approx 7$ T and  $B \approx 4$ T for $D=100\AA$ and
$D=50\AA$, respectively. This deviation points out the fact that inter-dot
correlations have grown to play their role, a role that cannot be described
in terms of a mean field theory.
A phase diagram is shown in the inset:
The inter-dot correlation grows to be relevant with the
inverse of the distance (as one would expect), and with the strength of
the magnetic field.
Now one can understand the appearance of the $\nu^{DQDS}=1$  from a different
point of view:
The new incompressible state at short distances ($M=15$) has
its origin in the spatial inter-dot correlations, and cannot be obtained as a
simple superposition of two equal QD stable electronic configurations.
The origin of such particular occupation
numbers can be understood. One can form a GS with $M=15$ out of two
single QD
configurations with $M^+=3$ and $M^-=12$. These two configurations match
perfectly with each other in the sense that the 3 magnetic flux quanta of the
$M^-=12$, i.e., the 3 quasiholes in the center of the lower QD "recombine"
with the 3 quasiparticles of the $M^+=3$ configuration of
the upper QD. Of
course, one cannot label the electrons of different dots due to
the inherent particle indistinguisibility so one must
think in terms of a linear combination of the above total configuration with
the reversed one: quasiholes in the upper dot and quasiparticles
in the lower one. Thus, the origin of the $\nu^{DQDS}=1$ from the spatial
inter-dot correlations becomes clear.

\section{conclusions}

In this work we have analyzed the new correlated electronic states appearing
in double quantum dot systems. The simplest case of two electrons was solved
analytically. By means of exact diagonalization of the full interacting
Hamiltonian for up to six electrons,
minimum-isospin ground states were found to appear for
short distances between identical quantum dots, and to
disappear as the distance increased. Mean field calculations revealed the
critical distances at which the inter-dot correlations  were significant,
signaling the appearance of the new minimum-isospin ground states.

\section*{acknowledgments}

One of the authors (JJP) wish to acknowledge the hospitality
and support of the Institute for Microstructural Sciences, NRC Canada,
where this work was conceived.
We are also indebted to L. Brey, A. H. MacDonald, and C. Tejedor for fruitful
discussions, and to G. Aers, and, again to L. Brey,
for the critical reading of the manuscript.
This work has been supported in part by the Comisi\'on
Interministerial
de Ciencia y Tecnolog\'{\i}a of Spain under contract
MAT 91-0201 and by the Commission of the European Communities under
contract No. SSC-CT-90-0020.

\begin{figure}
\caption{A schematic picture of a DQDS. Such type of semiconductor
structures can be now routinely fabricated by means of combined growing
and etching techniques.}
\label{fig1}
\end{figure}

\begin{figure}
\caption{Evolution of the energy spectrum as a function of the distance between
QD's of a 2-electron DQDS with total angular momentum $M=6$. Notice
the splitting of the $I=1$ set of states into $I_z=0$ and $I_z=\pm 1$ states
as the distance between QD's increases. The corresponding center of mass
and relative angular momenta
have been explicitly stated in the figure.}
\label{fig2}
\end{figure}

\begin{figure}
\caption{The phase diagram (angular momentum and isospin)
of the 2-electron DQDS ground states.}
\label{fig3}
\end{figure}

\begin{figure}
\caption{(a)Evolution of the absolute GS angular momentum $M$ as a function
of the single-particle kinetic energy (magnetic field/confinement) for
4 electrons in the ZDL. (b) The same for 6 electrons.
Along with $M$, the values
of the total isospin $I$ of the stable states discussed in the text
are also shown.}
\label{fig4}
\end{figure}

\begin{figure}
\caption{ The excitation spectrum of 6 electrons in a DQDS in the ZDL. The
GS is marked with a star. Isospin-reversed quasiparticle
excitations are distinguished
by solid dots and quasihole excitations by open dots. The arrows show those
subspaces whose GS's will become the absolute GS in the presence of isospin
(spin) Zeeman energy for increasing $B$ (to the right of $M=15$) and
for decreasing $B$ (to the left of $M=15$).}
\label{fig5}
\end{figure}

\begin{figure}
\caption{(a)Evolution of the absolute GS angular momentum $M$ as a function
of the single-particle kinetic energy (magnetic field/confinement) for
6 electrons for $D= 50 \AA$. (b) The same for  $D= 100 \AA$.}
\label{fig6}
\end{figure}

\begin{figure}
\caption{Single particle occupation numbers $\nu^+_m$ ($\nu^-_m$) of the
$\nu^{DQDS}=1$ state as a function of the distance.}
\label{fig7}
\end{figure}

\begin{figure}
\caption{Evolution of the renormalized GS energy (substracting $N\epsilon_0$)
 as a function of $B$. Crosses
show the HF approximation results discussed in the text and dots those
from the exact diagonalization calculations. The inset shows a phase
diagram for the inter-dot correlation as a function of $B$ and $D$.}
\label{fig8}
\end{figure}


\begin{references}
\bibitem{pra1} One must bear in mind that
the definition of filling factor for a DL2DES is twice the usual
definition for a single layer.
\end{references}
\end{document}